\newtheorem{conjecture}{Conjecture}

\newcommand{\ket}[1]{|\,#1\,\rangle}
\newcommand{\bra}[1]{\langle\,#1\,|}
\newcommand{\braket}[2]{\langle\,#1\,|\,#2\,\rangle}
\newcommand{\tr}[0]{\mbox{tr}}

\documentclass[12pt]{article}   
\usepackage{fullpage} 
\usepackage{graphicx}
\title{When Worlds Collide: \\ Quantum Probability From Observer Selection?}
\author{Robin Hanson\thanks{Department of Economics, George Mason University, MSN 1D3, Carow Hall, Fairfax VA 22030 http://hanson.gmu.edu rhanson@gmu.edu} \\
}

\date{April 7, 2003.  \\ First version August 9, 2001.}

\begin{document}

\maketitle


\begin{quote}
In Everett's many worlds interpretation, quantum measurements are considered to be decoherence events.  If so, then inexact decoherence may allow large worlds to mangle the memory of observers in small worlds, creating a cutoff in observable world size.  Smaller world are mangled and so not observed.  If this cutoff is much closer to the median measure size than to the median world size, the distribution of outcomes seen in unmangled worlds follows the Born rule.  Thus deviations from exact decoherence can allow the Born rule to be derived via world counting, with a finite number of worlds and no new fundamental physics.   
\end{quote}



\section{INTRODUCTION}

Traditionally, quantum systems have been described as evolving according to two different rules.  A usual deterministic linear evolution rule is occasionally replaced by a stochastic quantum measurement rule, which eliminates all but one diagonal element from the density matrix.  Unfortunately, this stochastic rule is ambiguous in several ways.  

The many worlds interpretation of quantum mechanics was intended to settle these ambiguities, by showing that standard linear evolution could predict phenomena very similar to that predicted by the stochastic quantum measurement rule, with a minimum of additional assumptions$^{(1,2)}$.  This approach has born fruit lately, as research into ``decoherence" has shown how off-diagonal elements are often naturally and dramatically suppressed due to coupling with a large environment$^{(3,4,5)}$.  This allows us to settle several measurement ambiguities, by assuming that the timing and observables of quantum measurements coincide with the timing and observables of decoherence.

Unfortunately, the many worlds interpretation still suffers from the problem that the straightforward way to calculate the probabilities of outcomes, i.e., counting the fraction of worlds with a given outcome, does not produce the standard Born rule for measurement probabilities$^{(6,7)}$.  Everett originally tried to deal with this by showing that in the limit of an infinite number of measurements the total measure of worlds that do not observe Born rule frequencies approaches zero$^{(1)}$.  Yet we never actually have an infinite number of measurements.  With a finite number of measurements, the measure of non-Born rule worlds remains finite.  While non-Born rule worlds might tend to be smaller, it is not clear why we should discount the observations made in such worlds.  After all, if worlds split according to standard ideal measurement processes, the future evolution of each world is entirely independent of the evolution of other worlds.

Some have tried to reconcile many worlds with the Born rule by adding new fundamental physics.  For example, some suggest adding non-linearities to state-vector dynamics$^{(8)}$.  Others postulate that an infinite number of worlds correspond to each possible outcome, and that some new physics makes the proportion of those worlds seeing each outcome follow the Born rule$^{(9)}$.  Still others propose that we accept decision theory axioms stating that we do not care about the number of worlds that see an outcome, and so our subjective probabilities need not reflect such numbers$^{(10,11)}$. 

In this paper I suggest how one can reconcile a many worlds approach with the Born rule, without introducing new fundamental physics and without changing decision theory.  That is, I suggest how one can derive the prediction that observed long run measurement frequencies are given by the Born rule, while still insisting that only the standard linear evolution rule always and exactly determines evolution, while calculating all predicted frequencies by counting world fractions, and while always considering only a finite number of worlds and measurements.  Of course some assumptions are required to get this result.  But these assumptions are about the behavior of standard linear quantum evolution, assumptions that can in principle be checked via more careful theoretical analysis of how standard linear evolution plays out in common quantum systems.  

The assumptions I make embody two main ideas.  
The first idea is that since decoherence is never exact, worlds do not exactly split according to an ideal measurement process.  The evolution of the density matrix term describing a world is influenced both by internal autonomous dynamics, \emph{and} by cross-world influences from off-diagonal density matrix terms.  While decoherence makes these off-diagonal terms small relative to a large world, they can be large relative to a small enough world.  So while to a good approximation large worlds evolve autonomously, the evolution of small worlds can be dominated by influences from larger worlds.  This may plausibly mangle small worlds, either destroying the observers in such worlds, or changing them into observers who remember events from large world.  (I also assume that mangling is more of a sudden than a gradual process.)

The second main idea is that in typical situations where we test the Born rule by counting the outcomes for some decoherence events, there are many more background events that we do not count.  Given many independent decoherence events that change world sizes (i.e., measures) by multiplicative factors, world sizes are distributed lognormally with a large variance.  In such a distribution, the median of the distribution of measure is high into the upper tail of the distribution of worlds, at a place where the world distribution happens to fall off as an inverse square.  Since it is the measure of some worlds that mangles other worlds, this median measure position is where we might expect to find a ``mangling region" in world size.  Worlds that go from being larger than this mangling region to smaller than it become mangled in the process.  This mangling process results in a transition region, where the fraction of worlds that are mangled goes from nearly zero at the top of the region, to nearly one at the bottom of the region. 

When a lognormal distribution over some value is truncated at such a median value point, dropping all smaller items, it so happens that multiplying the number of items in the distribution by a factor has the same effect as increasing the value of each item in the distribution by that same factor.   But this implies that if the entire transition region is near the median measure point, then the number of unmangled worlds corresponding to each outcome of a measurement is proportional to the total measure associated with that outcome, just as the Born rule requires.  Thus unmangled observers in almost all worlds would remember having observed frequencies near that predicted by the Born rule, even though in fact Born frequencies do not apply to the vast majority of worlds, and even though such frequencies are not observed in the very largest worlds.  A ``mangled worlds" variation on the many worlds interpretation may thus predict the Born rule for quantum probabilities.  

This mangled worlds approach may be experimentally testable.  Not only can small testable deviations from the exact Born rule arise, but this approach also seems to predict that physical states in which decoherence events are less frequent will predominate.    

In the sections that follow, I first review the basics of quantum measurement, decoherence, and the many worlds interpretation.  Then after discussing the implications of inexact decoherence for the autonomy of small world evolution and the distribution of world sizes, I show that the Born rule is satisfied under the assumptions given.  Along the way, I discuss whether mangling is local or global, reversible or irreversible, and the potentially problematic decoherence event rate prediction. 

\section{QUANTUM MEASUREMENT}

Quantum mechanics has traditionally described systems via unit-magnitude Hilbert space
vectors $\ket{\psi}$ which evolve according to two different rules.  Usually, vectors evolve deterministically according to the standard linear rule

\begin{equation} i \hbar \frac{d}{dt} \ket{\psi} = H \ket{\psi} , \end{equation}
but on occasion they instead evolve non-deterministically according to
\begin{equation} \ket{\psi} = \sum_a \ \ket{a} \braket{a}{\psi} \mbox{~becomes~} \ket{a} 
   \mbox{~with probability~} |\braket{a}{\psi}|^2  .  \end{equation}
This second process is said to correspond to the measurement of a value
$a$ for some ``observable," where the $\ket{a}$ and $a$ are respectively orthonormal eigenvectors and eigenvalues of that observable's Hermitian operator $A$.  More precisely, given a complete set of orthogonal projection operators $\{P_a\}_a$ (so $\sum_a P_a = 1$ and $P_a P_b = \delta_{ab} P_a$), a measurement on $\ket{\psi}$ produces a normalized $P_a \ket{\psi}$ with probability equal to the measure, i.e., amplitude-squared, of $P_a \ket{\psi}$.  

In place of a Hilbert space vector $\ket{\psi}$, one can equivalently describe a quantum 
system in terms of a unit-trace Hermitian ``density matrix" $\rho$, which usually evolves according to 

\begin{equation}  i \hbar \frac{d}{dt} \rho = H \rho - \rho H , 
\label{rhoevolve} \end{equation}
but on occasion instead evolves according to
\begin{equation}
  \rho \mbox{~becomes~a~normalized~} P_a \rho P_a^{} 
   \mbox{~with probability~} \tr(P_a \rho P_a)  , 
\label{rhomeasure}
\end{equation}
where again the outcome $P_a \rho P^{}_a$ coincides with the observation of a value $a$ for observable $A = \sum_a a P_a$.

For seventy five years, people have wrestled with ambiguities in the above description of quantum dynamics.  What determines the time $t$ and projections $\{P_a\}_a$ for each measurement?  And to what extent do $\ket{\psi}$ and $\rho$ describe the system itself, as opposed to our knowledge of the system?   

\section{DECOHERENCE}

Recently, great progress has been made in reducing the time and observable ambiguities of measurement.  In the density matrix formulation of quantum dynamics, a key distinguishing feature of the measurement process described in equation~\ref{rhomeasure} is that it displays ``decoherence."  That is, measurement eliminates the off-diagonal elements of $\rho$ in the $A$ representation of $\rho$.  Recently, detailed analyzes of many specific physical systems have shown that such decoherence usually comes naturally when a system interacts with a large environment via standard linear quantum evolution.   

The usual decoherence scenario describes a total system $\ket{\psi}\ket{E}$, which is a particular quantum system of interest $\ket{\psi}$ coupled to a large environment $\ket{E}$.  This scenario considers not the total density matrix $\rho_T$, but only the part of that matrix that describes the system $\ket{\psi}$.   That is, even though linear quantum evolution must preserve the off-diagonal elements of the total density matrix $\rho_T$, it need not preserve the off-diagonal elements of the system density matrix 
\begin{equation}  \rho = \tr_E( \rho_T ) , \end{equation}
where $\tr_E$ denotes a trace across the environment subspace.  Detailed analyzes of many specific situations have shown that initial system coherence in $\rho$ is usually transfered very quickly to the environment.  That is, the physics of such situations usually chooses particular projections $\{P_a\}$ for the system and induces a measurement-like evolution

\begin{equation}
   \rho = \sum_{ab} \rho_{ab}  \mbox{~~} \mbox{becomes} 
   \mbox{~~}  \approx \mbox{~~}  
   \rho' = \sum_{a} \rho_{aa}' , 
\label{decohere}
\end{equation}
where $\rho_{ab} \equiv P_a \rho P_b^{}$.  It seems that once coherence has been transfered from a quantum system to a large environment, it becomes for all practical purposes impossible to measure.  Because of this, we can eliminate some of the ambiguities of quantum measurement theory if we assume that the measurement process of equation~\ref{rhomeasure} only happens after decoherence of the same projections $P_a$ happen via equation~\ref{decohere}.

Of course even if we make this assumption, there still remains the ambiguity of the extent to which state vectors describe reality, as opposed to our knowledge of reality.  Furthermore, the measurement process of equation~\ref{rhomeasure} is still distinguished from the decoherence process of equation~\ref{decohere} by the fact that while decoherence eliminates the off-diagonal elements of $\rho$ in the $\{P_a\}_a$ representation, measurement also eliminates all but one of the diagonal elements of $\rho$.  And it does this non-deterministically. 

\section{MANY WORLDS INTERPRETATION}

Forty five years ago Hugh Everett proposed the ``many worlds," or ``relative state," interpretation of quantum mechanics in an attempt to resolve ambiguities of quantum measurement$^{(1,2)}$.  The many worlds interpretation posits that vectors $\ket{\psi}$ are literally real, and that they describe all physical systems, including human observers.  It posits furthermore that systems only evolve deterministically according to the standard linear quantum rule.  

Before an ideal many-worlds measurement, a total system vector is a product of an observed system vector $\ket{\psi}$ and an observer system vector $\ket{O}$.  During an ideal measurement, the total system evolves deterministically according to

\begin{equation}
  \ket{\psi} \ket{O} = \left( \sum_a \ket{a} \braket{a}{\psi} \right) \ket{O} \mbox{~~becomes~~} 
     \sum_a  \braket{a}{\psi} \ket{a} \ket{O_a} , 
\label{idealsplit}
\end{equation}
where $\ket{O_a}$ describes an observer who has observed the value $a$.  The apparent non-determinism of measurement evolution is resolved by positing that each $\ket{a} \ket{O_a}$ describes a different ``world," wherein a different copy of the original observer measured a different value of $a$.

The many worlds interpretation in essence posits that a careful analysis of standard linear deterministic quantum evolution will show that it reproduces all of the phenomena usually explained by invoking non-deterministic quantum evolution.  While Everett's original formulation of many-worlds did not specify when and along what projection basis worlds split, the recent decoherence analyzes have gone a long way to offering answers to such questions.  That is, it now seems plausible to say that the world of an observer splits when that observer observes a quantum system, and when both observer and observed are coupled to a large environment so as to produce decoherence across multiple relevant states of the observed quantum system.

There remains, however, one important measurement phenomena that the many worlds interpretation has not yet adequately accounted for.  That phenomena is the ``Born rule," i.e., the particular probability distribution according to which measurements produce states $\ket{a}$.  The Born rule states that after a measurement the vector $\ket{a}$ is seen with probability $|\braket{a}{\psi}|^2$, or that a normalized $\rho_{aa}$ is seen with probability given by its measure $\tr(\rho_{aa})$.  While this distribution can be derived from assumptions of unitary symmetry$^{(12)}$, a straightforward many-worlds observer-selection analysis gives a different answer.    

If we assume a finite number of worlds, and if we simply count the number of worlds in which observers see different long-run measurement frequencies, we do not find the Born rule$^{(6,7)}$.  We instead find that the vast majority of worlds display long-run frequencies consistent with a uniform distribution.  Under a uniform distribution, a normalized $\rho_{aa}$ is seen with probability $1/N_A$, where $N_A$ is the number of projections in the set $\{P_a\}_a$.  Such a distribution usually differs greatly from the Born rule.  

Everett tried to address this problem by showing that worlds containing long-run frequencies which differ substantially from the Born frequencies have a much smaller total measure than worlds containing frequencies very near the Born rule.  Specifically, Everett showed that in the limit of an infinite number of non-trivial measurements, the aggregate measure of non-Born frequency worlds approaches zero.  It is not clear, however, that there are ever an infinite number of non-trivial measurements.  With a finite number of measurements, observers who see non-Born rule frequencies might live in very small measure worlds, but it is not clear why such observers should not be counted.  After all, if world splittings happen according to the ideal measurement process described in equation~\ref{idealsplit}, the future evolution of a world should be entirely independent of the amplitude of that world, and so observers in such worlds should live out their lives just as observers in any other world would.

To resolve this problem, some have proposed that we accept decision theory axioms which state that we do not care about the number of worlds that see an outcome, and so our subjective decision probabilities need not reflect that number$^{(10,11)}$.  
Others have postulated that an infinite number of worlds correspond to each possible measured state, and that by some as yet unknown physical process these worlds diverge during a measurement in proportion to the Born rule measure$^{(9)}$.  Still others propose adding non-linearities to state-vector dynamics$^{(8)}$.  In this paper, we retain standard decision theory and the traditional focus on world counting, and we assume that both the number of worlds and the number of measurement-like events are finite.

\section{INEXACT DECOHERENCE}

We noted above that if world splittings happen according to the ideal measurement process described in equation~\ref{idealsplit}, the future evolution of a world should be entirely independent of the amplitude of that world.  In fact, however, real measurements are not usually exactly ideal.  And if the decoherence equation~\ref{decohere} is only approximate, then the measurement equation~\ref{idealsplit} is only approximate as well.  Small as these effects may seem, they may force a substantial revision to our conclusions about
observed long-run measurement frequencies in a many worlds interpretation.   

Let us consider a particular large world $\ket{L}$ and small world $\ket{s}$, and define their relative size $\delta(t)$ via

\begin{equation}   | \rho_{ss} |  \approx \delta(t)^2  | \rho_{LL} | ,
\label{bigsmall} \end{equation}
\noindent where magnitude is $|\rho_{aa}| \equiv \tr(\rho_{aa})$.  
Next let us consider the relative magnitude of diagonal terms $\rho_{aa}$ and off-diagonal terms $\rho_{ab}$ for $a \not= b$.  
Roughly following standard notation$^{(4)}$, we may describe a degree of decoherence $\epsilon(t)$ as

\begin{equation} | \rho_{ab} |^2 \leq  \epsilon^2(t)  | \rho_{aa} | | \rho_{bb} |   , \end{equation}
where for fine grain projections $P_a = \ket{a}\bra{a}$, we can define magnitude as 
$|\rho_{ab}| \equiv \bra{a}\rho_{ab}\ket{b}$.  (Similar results obtain if we substitute $(| \rho_{aa} | + | \rho_{bb} |)^2$ for 
$| \rho_{aa} | | \rho_{bb} |$ in the definition of $\epsilon(t)$.  It is not clear to me how best to define magnitude $|\rho_{ab}|$ for coarser projections.)  

Let us therefore write 

\begin{equation} |\rho_{Ls}|^2 \approx |\rho_{sL}|^2 \approx  \epsilon^2(t)  |\rho_{LL}| |\rho_{ss}|  . \end{equation}
Combining this with equation~\ref{bigsmall} gives these relative magnitudes
\begin{eqnarray}
     |\rho_{LL}|  &  &  \approx \mbox{~~} 1,  \\
     |\rho_{Ls}|  & \approx \mbox{~~} |\rho_{sL}| & \approx \mbox{~~} \epsilon \delta, \\
     |\rho_{ss}|  &  &  \approx \mbox{~~} \delta^2 . 
\end{eqnarray}
Note that even when coherence $\epsilon$ is very small, if the relative size $\delta$ is smaller still, then off-diagonal terms like $\rho_{Ls}$ can still have larger magnitudes than small world diagonal terms like $\rho_{ss}$.

How small does coherence $\epsilon$ get?  Detailed analyzes of many specific situations have found that while the coherence $\epsilon(t)$ typically falls at a rapid exponential rate for a great many doubling times, it eventually asymptotes to a small but non-zero level.  For example, this happens in several models where a particular observable of a single simple particle is continually ``measured" by an infinite (or finite) environment, and where this measurement continues on forever$^{(3,4,13,14)}$. 

I know of no detailed analyzes of models where the same environment repeatedly measures different non-commuting observables.  Let us conjecture, however, that such models will also find that coherence eventually either asymptotes to a positive value, or at least falls at a slower than exponential rate:

\begin{conjecture}
After two worlds split due to a decoherence event, their coherence $\epsilon(t)$ typically falls with time $t$, but eventually falls slower than $e^{\sqrt{rt}}$, where $r$ is the effective rate of further decoherence events.
\label{coherencelarger}
\end{conjecture}

\section{WORLDS COLLIDING}

Consider the co-evolution of a large world $\ket{L}$ and a small world $\ket{s}$, which have almost but not exactly decohered due to coupling with some environment.  The total system of worlds plus environment must evolve according to the standard linear rule of equation~\ref{rhoevolve}, but the partial system of the worlds alone need not.  If we treat the interaction with the environment as weak we can write
\begin{equation}  i \hbar \frac{d}{dt} \rho = H \rho - \rho H + S \end{equation}
where $H$ describes the evolution of isolated worlds, and $S$ describes the 
change in $\rho$ due to interaction with the environment.
If we decompose this according to the two projections $P_{L}$ and $P_{s}$ for the two worlds, we get 

\begin{eqnarray}
  i \hbar \frac{d}{dt} \rho_{LL} =&  H_{LL} \, \rho_{LL} -  \rho_{LL} \, H_{LL}  
                                    &+ (H_{Ls} \, \rho_{sL} -  \rho_{Ls} \, H_{sL}) + S_{LL} 
\label{LL}  \\
  i \hbar \frac{d}{dt} \rho_{ss} =&  H_{ss} \, \rho_{ss} -  \rho_{ss} \, H_{ss} 
                                    &+ (H_{Ls} \, \rho_{sL} -  \rho_{Ls} \, H_{sL}) + S_{ss} 
\label{ss}  \\
  i \hbar \frac{d}{dt} \rho_{Ls} =&  H_{LL} \, \rho_{Ls} -  \rho_{LL} \, H_{Ls}  
                                    &+ H_{Ls} \, \rho_{ss} -  \rho_{Ls} \, H_{ss} + S_{Ls}  
\label{Ls}  \\
  i \hbar \frac{d}{dt} \rho_{sL} =&  H_{sL} \, \rho_{LL} -  \rho_{sL} \, H_{LL}  
                                    &+ H_{ss} \, \rho_{sL} -  \rho_{ss} \, H_{sL} + S_{sL}   
\label{sL} 
\end{eqnarray}
Equations~\ref{LL}~and~\ref{ss} describe the evolution of the two worlds, and the terms in
parentheses there describe evolution due to influence from off-diagonal terms.  Equations~\ref{Ls}~and~\ref{sL} describe the evolution of the off-diagonal terms, in part due to influence from the two worlds.

Assume that the various $H$ terms are of similar magnitudes, and that the relative magnitude of terms roughly determines the relative strength of influence of those terms.  If so, then the autonomy of a world's evolution depends primarily on the relative magnitudes of the various density matrix terms.  We have assumed that $|\rho_{ss}| \approx \delta^2 |\rho_{LL}|$ and $|\rho_{sL}| \approx |\rho_{Ls}| \approx \epsilon \delta |\rho_{LL}|$.  This implies that (for $\delta,\epsilon \ll 1$ and  $S$ small) $\rho_{LL}$ is by far the largest influence on the evolution of $\rho_{LL}$.  That is, to a good approximation the large world evolves autonomously.  In contrast, for $\delta < \epsilon \ll 1$ the evolution of $\rho_{ss}$ is determined at least as much by $\rho_{sL}$ and $\rho_{Ls}$ as by $\rho_{ss}$.  And the evolution of $\rho_{sL}$ and $\rho_{Ls}$ is dominated by $\rho_{LL}$.  That is, in a situation like this the evolution of the small world is mostly slaved to the evolution of the large world, via the off-diagonal intermediaries.  

If, as we have assumed, $P_s$ and $P_L$ are the projections that make the two worlds seem the most decoherent, then we simply cannot consider the small world $s$ to be evolving autonomously as suggested by the idealized many worlds measurement of equation~\ref{idealsplit}, since the evolution of typical measurement records and observers in the small world $s$ will be determined primarily by influences from measurement records and observers in the large world $L$.  Such strong influence seems likely to either destroy such small world measurement records and observers, since as physical systems they were not designed to deal with such perturbations, or to change those small world measurement records and observers into ones like those found in the large world.  This suggests another conjecture.

\begin{conjecture}
When the coherence $\epsilon$ between two worlds is large enough compared to their relative measure $\delta$, human observers in the small world will typically be ``mangled," i.e., will either fail to exist or will remember the measurement frequency of the large world.  
\label{nosmallobserver} 
\end{conjecture}

Note that this conjecture need only apply to typical current human observers.  Perhaps, by using quantum error correction codes, quantum computers and the humans that observe them would be better able to resist the influence of larger worlds via off diagonal terms.  

Combining conjectures~\ref{coherencelarger} and \ref{nosmallobserver}, we can see that if the relative magnitude between two worlds grows exponentially with time (or at least grows as $e^{\sqrt{rt}}$), then the coherence between these worlds will eventually become large compared to their relative magnitude, and so the observers in the small world will become mangled due to interactions between these worlds.   So when does the relative magnitude of worlds grow exponentially?  

\section{THE DISTRIBUTION OF WORLD SIZES}

Consider a single initial world that repeatedly undergoes decoherence events $e$.  During each event $e$, each pre-existing world $i$ splits into a set $J(i,e)$ of resulting worlds $j$, each of which gets some fraction $F_{j i e}$ of the original world's measure.   That is, if $m_i$ is the measure (or size) of world $i$, then $m_j = F_{j i e} m_i$, where\footnote{Inexact decoherence may not allow this next formula to hold exactly.} $\sum_{j \in J(i,e)} F_{j i e} = 1$.  Let us assume, as measurement analyzes commonly do, that the fractions do not depend on the particular world being split, so that $F_{j i e} = F_{j e}$ and $J(i,e) = J(e)$.  If so, then after there have been enough events so that the effect of each event is relatively small\footnote{Technically, the variance of each set
$\{\log(F_{je})\}_{j \in J(e)}$ should be small compared to this summed over $e$.}, the central limit theorem of statistics ensures that the resulting set of worlds will, to a good approximation, be distributed log-normally over measure $m$.  

That is, a single unit measure world would give rise to a set of worlds normally distributed in $\log(m)$, with some mean $\log(\tilde{m}) < 0$ and standard deviation $\sigma > 0$.  The world of size $\tilde{m}$ would also be the median sized world, the world where half of all worlds are larger, and half are smaller.  The total number of such worlds would be about $1/\tilde{m} e^{\sigma^2/2}$.  The measure held by these worlds would also be distributed normally over $\log(m)$, with the same standard deviation $\sigma$, but a much higher mean and median of $\log(\hat{m}) = \log(\tilde{m}) + \sigma^2$.  That is, worlds with sizes within a few $\sigma$ of $\log(\hat{m})$ would contain almost all of the measure.  For large $\sigma$, these would be a very small fraction of the total number of worlds, though a very large number of worlds.  
Let us conjecture that world size is distributed lognormally with a large $\sigma$, mostly due to background decoherence events.

\begin{conjecture}
Typical situations where we test the Born rule are the result of a large number ($> 10^4$) of mostly uncounted decoherence events, each of which has a small fractional influence.  Even when we only count frequencies from a few events, many other background events occur.   Thus the distribution of world size is lognormal, with $\sigma$ large ($> 50$).
\label{lognormal} 
\end{conjecture}

For example, consider a system that undergoes a set of binary decoherence events, where each event has two possible outcomes, with relative measure and Born rule probabilities $p > 1/2$ for ``up" and $1-p$ for ``down."  After $N \gg 1$ such events, measure would be normally distributed over log world size with a mean and median of

\begin{equation} \log(\hat{m}) = N \log(\hat{m}_1) = N(p\log(p)+(1-p)\log(1-p)) \end{equation}
and a standard deviation of 

\begin{equation} \sigma = \sqrt{N} \sigma_1 = \sqrt{N}  \sqrt{p(1-p)} \log(\frac{p}{1-p}) .\end{equation}
The worlds themselves would also be distributed normally in log size, with the same standard deviation but a lower median; the median world would be of size $\tilde{m} = \hat{m}e^{-\sigma^2}$.  For example, $10^4$ binary decoherence events, at $p=.75$, would produce $2^{10^4}$ worlds with a standard deviation of $\sigma \approx 50$ (in $\log(m)$ units).

(Note that the very largest worlds would have seen almost no ``down" events, the very smallest worlds would have seen almost no ``up" events, and the vast majority of worlds would see nearly equal numbers of ``up" and ``down" events.  So I have not yet reproduced the Born rule.  But bear with me.)

If decoherence events occured at a rate $r$, so that $N = rt$, then the size of the median world would fall exponentially, as

\begin{equation} \tilde{m}(t) = (\hat{m}_1 e^{-\sigma^2_1})^{rt} , \end{equation}
and the typical relative magnitude between worlds would grow according to

\begin{equation} e^\sigma(t) = (e^{\sigma_1})^{\sqrt{rt}} . \end{equation}  

\section{THE MANGLING TRANSITION}

The above growth rate in relative size, together with conjectures~\ref{coherencelarger} and \ref{nosmallobserver}, should ensure that for two randomly chosen worlds, typically one of them eventually mangles the other.  
Given the enormous number of worlds involved, this suggests that smaller than normal worlds are almost sure to be eventually mangled.  Also, the very largest worlds cannot be mangled, because there are no worlds large enough to have done the job.  There must therefore be some transition region $[\underline{m},\overline{m}]$ in world size.  Below the bottom of this region $\underline{m}$, worlds are almost surely mangled, while above top of this region $\overline{m}$, worlds are almost surely not mangled.  Within the transition region, the fraction of mangled worlds should gradually fall as one moves from the bottom of this region to the top.\footnote{Note that the transition region need not be the same as the mangling region, the region in which unmangled worlds become mangled.}

What can we say about the size and location of this transition region?  If smaller than normal worlds are almost surely mangled, the bottom of this region must lie well above the median world size, so that $\underline{m} \gg \tilde{m}$.  There are also many other assumptions we might plausibly make.  For example, we might assume that if any substantial fraction of worlds of size $m$ get mangled, virtually all worlds of size $e^{-50}m$ would also be mangled, which implies $\log(\overline{m}/\underline{m}) < 50$.  And since it is fundamentally the measure of some worlds that does the mangling of other worlds, we might conjecture that the transition region lies within a few $\sigma$ of the median measure $\hat{m}$.  

While many such assumptions seem plausible, let us here only assume what we need for the analysis that follows. It turns out to be sufficient to assume that all the world sizes in the transition region lie much closer to the size of the world at the median measure position $\hat{m}$ than to the world with the median size $\tilde{m} = \hat{m}e^{- \sigma^2}$.  

\begin{conjecture}
In typical situations where we test the Born rule by counting frequencies, there are far more uncounted background events.   Decoherence events are mostly independent and fractionally small, resulting in a lognormal distribution over world size.  There is an outcome independent transition region $[\underline{m},\overline{m}]$ in world size $m$, below which worlds are mangled and above which they are not.  For all $m \in [\underline{m},\overline{m}]$, we have $| \log(m /\hat{m}) | \ll \sigma^2 \gg 1$.
\label{cutoff} 
\end{conjecture}

The outcome independence assumption says that the fraction of worlds of a given size that are mangled does not depend on the experimental outcomes they observe.  Note that while conjecture~\ref{cutoff} requires that the vast majority of worlds be mangled, it is consistent with the vast majority of measure being either in mangled worlds (via $\underline{m} \gg \hat{m}$), or in unmangled worlds (via $\overline{m} \ll \hat{m}$).  

To see the implications of this conjecture, let $A(m)$ be the number of worlds with size less than $m$, so that $D(m) = A'(m)$ is the (lognormally distributed) density of worlds.  We can then define a local-power (or elasticity) as 

\begin{equation} \alpha(m) = \frac{d\log(D(m))}{d\log(m)} = \frac{\log(\tilde{m}/m)}{\sigma^2} - 1 = \frac{\log(\hat{m}/m)}{\sigma^2} - 2   . \end{equation}
It is reasonable to call this a local-power because $D(m)$ goes as (i.e., is proportional to) $m^\alpha$ in the neighborhood of $m$.  For example, $\alpha(\tilde{m}) = -1$, which says that a lognormal distribution $D(m)$ goes as $1/m$ near its median $\tilde{m}$.  We also have $\alpha(\hat{m}) = -2$, which says that $D(m)$ goes as $m^{-2}$ near the median measure $\hat{m}$.  

By assuming that the transition region is near the median measure $\hat{m}$, conjecture~\ref{cutoff} in essence assumes that $|\alpha({m})+2| \ll 1$ for $m \in [\underline{m},\overline{m}]$.  That is, we have assumed that $D(m)$ goes very nearly as $m^{-2}$ throughout and nearby the transition region.  

A distribution that goes as $m^{-2}$ falls off very rapidly with increasing $m$.  So one implication of $D(m)$ going as $m^{-2}$ at the transition region is that the vast majority of non-mangled worlds lie within or just above the transition region.  If the transition region was a zero width cutoff, the vast majority of non-mangled worlds would be within the few $\log(m)$ units above this cutoff.  And if the transition region were at least a few $\log(m)$ units wide, then the vast majority of non-mangled worlds would lie within the transition region itself. 

One conclusion we might draw from this implication is that world mangling is likely to mainly be a sudden, rather than a gradual, process.  After all, if mangling were gradual, so that worlds gradually became more mangled as they became smaller, and if the transition and mangling regions were nearly the same, then since most worlds are near or in the transition region, most worlds would be partially mangled.  This would suggest that we should see evidence of our world being partially mangled.  The fact that we do not notice evidence of such partial mangling suggests that mangling is largely a sudden process, appropriately described as ``collisions" between worlds.

\section{OBSERVED FREQUENCIES}
\label{split}

Conjecture~\ref{cutoff} also implies that the measurement outcomes that are associated with the largest total measure are the outcomes associated with the largest number of unmangled worlds.  This means that near Born rule frequencies will be observed in most unmangled worlds.  (The previous conjectures are not needed directly; they served to motivate conjecture~\ref{cutoff}.)

To see why this is so, consider a set of worlds which all undergo one or more decoherence events, some of which correspond to measurements.  Each initial parent world would be split into a finite set of child worlds which vary in their measurement outcomes.  If these decoherence events treated each world in the same way, then each distinct measurement outcome $k$ would be associated with a set of $G_k > 1 $ child worlds per parent world, and the measure of each child world would be a factor $F_k < 1$ smaller than the measure of its parent.  (And we should have $\sum_k F_k G_k = 1$.)

Given an initial distribution of worlds $D(m) = A'(m)$, there would be a new distribution of worlds $D_k(m) = A'_k(m)$ associated with each particular measurement outcome $k$.  Since when the mangling transition region is near $\hat{m}$, the vast majority of unmangled worlds lie in or just above the transition region, the number of worlds that observe a particular outcome $k$ is basically given by the value of $D_k(m)$ in and perhaps just above the transition region.

Imagine that there were some unmangled-fraction function $\gamma(m)$, where $\gamma(\overline{m}) \approx 1$, $\gamma(\underline{m}) \approx 0$, and $\gamma'(m) > 0$.  The total number of unmangled worlds for outcome $k$ would then be given by 

\begin{equation} \int_{-\infty}^\infty \gamma(m) D_k(m) dm \approx \int_{\underline{m}}^{\overline{m}} \gamma(m) D_k(m) dm + 
                                             \int_{\overline{m}}^{\infty} D_k(m) dm  . \end{equation}
As long as the different outcomes share the same mangle-fraction $\gamma(m)$, an outcome that has a larger $D_k(m)$ over and just above the region $[\underline{m},\overline{m}]$ will have more unmangled worlds.\footnote{If $\gamma(m)$ is large enough just above $\underline{m}$, a larger $D_k(m)$ in this region suffices.  This is probably the typical case.}  

In general, each outcome distribution $D_k(m)$ is given by 

\begin{equation} dA_k = D_k(F_k m) d(F_k m) = G_k D(m) dm , \end{equation}
which solves to 

\begin{equation}  D_k(m) =  G_k D(m/F_k)/ F_k . \end{equation}
Since for sizes $m$ near $\hat{m}$ density $D(m)$ goes nearly as $m^{-2}$, this implies that 
\begin{equation}  D_k(m) \approx F_k G_k D(m) \end{equation}
for $m$ near $\hat{m}$.  Thus if both $\underline{m}$ and $\overline{m}$ are near enough to $\hat{m}$, the number of worlds that see the measurement outcome $k$ is proportional to the product $F_k G_k$, which is exactly the fraction of the initial measure that is associated with that outcome.\footnote{Note that the converse also holds.  When $\underline{m}$ and $\overline{m}$ are not near $\hat{m}$, density $D(m)$ does not go nearly as $m^{-2}$, and so the number of unmangled worlds is not proportional to $F_k G_k$.  In general, many worlds models need not reproduce the Born rule.}

Thus given the assumptions we have made, the number of worlds which see a particular measurement outcome becomes proportional to the measure of the worlds that see that measurement outcome.  And since the Born rule frequencies come from weighting outcomes by their measure, this means that our assumptions predict that the vast majority of unmangled worlds will observe near Born rules frequencies.  

For example, consider the case of $N$ independent binary measurements, each with a measure (and probability) of $p$ for one outcome (e.g., ``up") and $1-p$ for the other outcome (e.g., ``down").  There are $N+1$ possible frequencies $f=M/N$ that can be observed, where $M$ is an integer in $[0,N]$.  The count $C$ of worlds at each frequency $f$ is given by

\begin{equation} C(f) = \frac{N!}{M! (N-M)!}  ,  \end{equation}
and the relative measure $m$ of each such world is given by

\begin{equation}m(f) = p^{M} (1-p)^{N-M} = (p^f (1-p)^{1-f})^N . \end{equation}
Since this last relation makes $\log(m)$ linear in $f$, the world count function $C(f)$ is proportional to $m D(m)$, which is the density of worlds in $\log(m)$ units (i.e., the number of worlds whose size falls in a small interval $d \log(m)$).  

\begin{figure}
\includegraphics[width=3.75in,height=5.75in,angle=270]{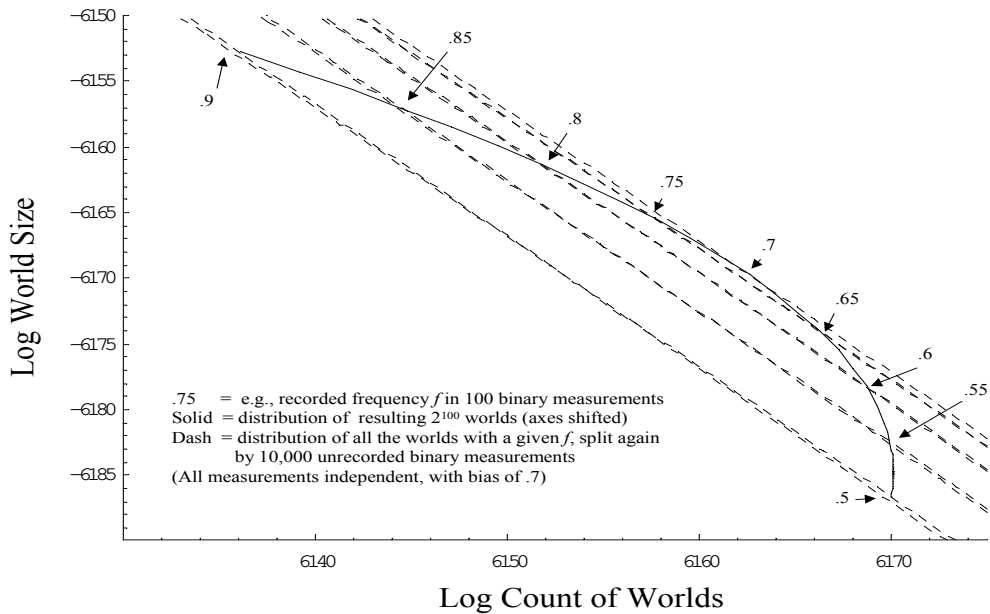}
\caption{The Distribution of Worlds} 
\label{worlds}
\end{figure}


The solid line in figure~\ref{worlds} shows how the ($e$-based log of) world count $C(f)$ and world size $m(f)$ vary as the observed frequency is varied in $f \in [.5,.9]$ for the case of $100$ independent binary measurements and $p=.7$.  
(The numbers on the axes apply to the dashed lines, not the solid lines.  The solid line has been offset to allow it to be compared to the dashed lines.)  Note that the vast majority of worlds are to be found near the uniform distribution frequency $f=1/2$, that the few largest worlds are near $f=1$, and that frequencies near $f=p=.7$ have the largest values of $\log(C) + \log(m)$, and hence of total measure $C(f)m(f)$.   

Now imagine that these 100 binary events are counted in an experimental test of the Born rule, but that during this test an additional $10,000$ decoherence events occur in the background.  To model this, we shall split each world described in the solid line of figure~\ref{worlds} via an additional $10,000$ independent binary measurement-like events (also with $p=.7$ for simplicity).  Let us continue to label each world with the frequency $f$ it displayed in those first 100 events, the ones counted in the experiment, and let $f'$ be the frequency in the additional background events.  In this case, for each frequency $f$ there will be an entire distribution of worlds of varying sizes corresponding to the different background $f'$.

Each of the nine dashed lines in figure~\ref{worlds} shows the distribution of world sizes corresponding to one of the nine frequencies $f=.5,.55,.6,.65,.7,.75,.8,.85,.9$, marked where that dashed line crosses the solid line.  (Specifically, it shows the relation between $C(f')C(f)$ and $m(f')m(f)$ as $f'$ is varied.)  As one can see, within the range shown in the graph, the world distributions for frequencies $f$ that are closer to the Born frequency of $f=p=.7$ dominate the distributions for frequencies that are further from $f=p=.7$.  (One distribution dominates another if it has a higher count at every size.)  

This domination continues to hold over a wide range of world sizes.  For example, while exact Born frequency worlds have a log size of $-6170$, you have to go up to $-5956$ to see the lines for the frequencies $.75$ and $.7$ cross, and down to $-6384$ to see the lines for frequencies $.65$ and $.7$ cross.  So in this example a transition anywhere in a range of a factor of $10^{185}$ would still ensure that most unmangled worlds would observe a frequency within the range $[.65,.75]$ for the 100 events counted in the measurement .  

\section{GLOBAL VERSUS LOCAL MANGLING}

We have seen how a mangled worlds variation on pure linear many-worlds quantum mechanics can predict the Born probability rule.  When the transition region is much closer to the median measure world size $\hat{m}$ than to the median world size $\tilde{m}$, observers in the vast majority of unmangled worlds will observe near Born frequencies.

If the transition region is not exactly at the median measure size $\hat{m}$, however, then on average unmangled worlds should not expect to see exact Born frequencies.  There is thus a possibility that this theory could be tested by looking for slight deviations from Born frequencies in quantum measurements.  It also raises the question of how close we should expect the transition region to be to the median measure.

The size of the discrepancy we expect from the exact Born probability rule depends on the magnitude of the deviation

\begin{equation} \alpha(m) - 2 = \frac{\log(m /\hat{m})}{\sigma^2}  \end{equation}
for $m \in [\underline{m},\overline{m}]$ in the transition region.  
This deviation should be very small if two assumptions are satisfied: if $\sigma$ is very large, and if the transition region is not terribly far out into the tails of the distribution of measure.  This second assumption leads us to expect a moderate value for the (standard normally distributed) ratio

\begin{equation} z(m) =  \frac{\log(m /\hat{m})}{\sigma}  \end{equation}
for $m \in [\underline{m},\overline{m}]$ in the transition region.  That is, we expect $|z(m)|$ to be not much more than 10 (or perhaps 100 or 1000).  Since the deviation at issue can be written as

\begin{equation}\alpha(m) - 2 = z(m)/\sigma , \end{equation}
a moderate value of $z(m)$ and a large value of $\sigma$ together imply a small value for the deviation $\alpha(m)-2$.  

Why might we expect $\sigma$ to be large?  One reason would be if the mangling process were largely global, rather than local.  Think of worlds as sitting in a parameter or phase space, with the child worlds of a given parent world sitting relatively close to each other in this phase space.  If worlds are far more likely to collide when they share a recent common ancestor, then mangling may be largely local, so that the relevant distribution and value of $\sigma$ is local, and the transition region $[\underline{m},\overline{m}]$ varies across the phase space.  On the other hand, if the chance of worlds colliding does not much depend on how recently they had a common ancestor, then mangling may be largely global, so that a global $\sigma$ is relevant, and the transition region $[\underline{m},\overline{m}]$ is a common global feature.  

If the mangling process were largely global, then the relevant distribution of worlds may be the result of decoherence events stretching back until the early universe.  This might be $N = 10^{10}$ or $10^{100}$ or even more decoherence events, which would clearly correspond to an enormous standard deviation $\sigma$.  

In general, to calculate the predicted probability of a certain outcome of an experiment, one would take the initial distribution of worlds, select the worlds that are consistent with the initial conditions of the experiment, use quantum mechanics to follow those worlds to the resulting distribution of descendant worlds at the end of the experiment, and then see what fraction of those descendant worlds that are unmangled also have the particular outcome of interest.  While the set of worlds that match the given initial conditions of the experiment may have a smaller standard deviation $\sigma$ than the full set, with global mangling it is still likely to be an enormous set, ensuring very close agreement with the Born probability rule.  

On the other hand, if the mangling process were largely local, then one would follow a similar procedure to make experimental predictions, except that the relevant distributions would be local, containing far fewer worlds and a much smaller $\sigma$.  There would then be more scope for finding predictions that deviate from the Born probability rule. 

\section{THE REVERSIBILITY OF MANGLING}

The process by which a large world mangles a small world might be typically reversible, so that the small world becomes unmangled if it again becomes large enough.  Alternatively, this process might typically be in practice irreversible, so that it was thermodynamically unlikely that a mangled world would become unmangled.  

The world transition region is the region in world size where the fraction of mangled worlds goes from nearly zero to nearly one.  The world mangling region, in contrast, is the region where worlds that go from being larger than this region to smaller than this region become mangled in the process.  The above analysis has been expressed in terms of properties of the transition region, not the mangling region.  What can we say about the relation between these two regions?

In order for these to regions to be the same, world mangling would have to be reversible.  After all, worlds are distributed lognormally in size because the relative size of worlds drift via random walks.  So if the mangling region width were much narrower than the range $\sigma$ over which worlds wander, then due to these wanderings worlds would not only move from above to below the mangling region, they would also move from below to above the mangling region.  If mangled worlds typically stayed mangled when they moved back above the mangling region, then the top of the transition region would have to be higher than the top of the mangling region.\footnote{I thank Michael Weissman for pointing this out.}  (An explicit drift-diffusion model of this case is worked out in a companion paper$^{(16)}$.)

Note that if mangling is reversible, then once mangled but now unmangled worlds would have to look as if they had never been mangled.  After all, the vast majority of unmangled worlds would have once been mangled, and yet we do not seem to have historical records of previous periods when our world was mangled.  So it seems that if mangling is largely reversible, the process of unmangling must eliminate any records of the mangling period.  

\section{THE PROBLEM OF VARYING DECOHERENCE RATES}

What happens if the rate at which decoherence events occur varies with observable outcomes.  Imagine two types of quantum outcomes, each associated with a rate $r = dN/dt$ of decoherence events, where $r_1 < r_2$.  That is, imagine that one outcome consistently produced physical systems where decoherence events were less frequent.  If so, then dependencies like $\tilde{m}(t) = (\hat{m}_1 e^{-\sigma^2_1})^{rt}$ would eventually lead to the low rate outcome having much larger worlds, via $\tilde{m}_1 - \tilde{m}_2 \gg \sigma$ and $\hat{m}_1 - \hat{m}_2 \gg \sigma$.  If mangling were global enough to allow these sets of worlds to mangle each other, with a common transition region $[\underline{m},\overline{m}]$, then eventually virtually all unmangled worlds would be associated with the slow decoherence rate $r_1$.  

The world mangling process thus seems to select physical states associated with low decoherence rates\footnote{I thank Michael Weissman for pointing this out.}.  The mangled worlds approach to quantum mechanics therefore makes the clear prediction that observed physical systems will have near the lowest possible decoherence rates.  If the rate of decoherence events is mostly due to couplings with distant environments, such as via cosmological photon, graviton, or other background radiation, it may be hard to find local, as opposed to cosmological, predictions.  Nevertheless, there are likely to be strong predictions of some sort.  Further work is required to identify these predictions, and compare them to what we have observed or can observe. 

\section{CONCLUSION}

This paper has suggested that, if certain assumptions hold, the many worlds interpretation may predict the Born rule of quantum measurement probabilities, via world counting over a finite number of worlds, and without introducing new fundamental physics or changing decision theory.  The basic idea is that although the coherence between different worlds typically falls very rapidly at first, it seems to eventually fall more slowly than the relative size of worlds increases.  If so, it seems that eventually typical human observers in small worlds become ``mangled," i.e., are either destroyed, or fail to remember the measurement frequencies of those small worlds.  There should thus be a transition region in world size, below which worlds are mangled and above which they are not.  

Besides the mangling of small worlds by large ones, the other big assumptions required are that this transition region lies much closer to the median measure world size than to the size of the median world, and that tests measuring Born rule frequencies do not count most decoherence events.  These conjectures should be open to confirmation or rejection by more detailed theoretical analysis of the evolution of specific quantum systems.  If these conjectures are confirmed, then a ``mangled worlds" variation on the many worlds interpretation can predict that observers in the vast majority of worlds will recall seeing near Born frequencies for quantum measurements.  The many worlds interpretation would then be a big step closer to a satisfactory account of the ambiguities in quantum measurement.  

\section{ACKNOWLEDGMENTS} 

For their comments, I thank an anonymous referee, Nick Bostrom, David Deutsch, Matthew Donald, Hal Finney, Lev Vaidman, and especially Michael Weissman, whose comments and encouragement made a huge difference. 

\section{REFERENCES} 

1. H.~Everett, `` `Relative state' formulation of quantum mechanics,"
 {\em Rev. Mod. Phys.}, {\bf 29}, 454-462 (1957).

2. B.S. DeWitt and N.~Graham,  {\em The Many-Worlds Interpretation of Quantum Mechanics} 
(Princeton University Press, Princeton, 1973).

3. M.~Namiki, S.~Pascazio, and H.~Nakazato, {\em Decoherence and Quantum Measurements}
 (World Scientific, Singapore, 1997).

4. H.F. Dowker and J.J. Halliwell, ``Quantum mechanics of history: The decoherence functional in quantum
  mechanics," {\em Phys. Rev. D}, {\bf 46(4)}, 1580-1609 (1992).

5. E.~Joos, ``Elements of environmental decoherence," In P.~Blanchard, D.~Giulini, E.~Joos, C.~Kiefer, and I.-O.
  Stamatescu, editors, {\em Decoherence: Theoretical, Experimental, and
  Conceptual Problems}, 1-17 (Springer, Berlin, 2000).

6. A.~Kent, ``Against many-worlds interpretations,"  {\em Int. J. Mod. Phys. A}, {\bf 5}, 1745-1762 (1990).

7. G.~Auletta, {\em Foundations and Interpretation of Quantum Mechanics} (World Scientific, Singapore, 2000).

8. M.B. Weissman, ``Emergent measure-dependent probabilities from modified quantum dynamics without state-vector reduction,"  {\em Found. Phys.}, {\bf 12}, 407-426 (1999).

9. D.~Albert and B.~Loewer, ``Interpreting the many-worlds interpretation," {\it Synthese}, {\bf 77}, 195-213 (1988).

10. D.~Deutsch, ``Quantum theory of probability and decisions," {\em Proc. R. Soc. London, Ser. A}, {\bf 455}, 3129-3197 (1999).

11. D.~Wallace, ``Quantum probability and decision theory, revisited," quant-ph/0211104 (2002).

12. A.M.~Gleason, ``Measures on the closed subspaces of a {H}ilbert space,"  {\em J. Math. Mechanics}, {\bf 6},  885-894 (1957).

13. H.F. Dowker and J.J. Halliwell,  ``The decoherence functional in the caldeira-leggett model," In J.J. Halliwell, J.~Perez-Mercader, and W.H. Zurek, editors, {\em
  Physical Origins of Time Asymmetry}, 234-245 (Cambridge University
  Press, Cambridge, 1994).

14. W.G. Unruh and W.H. Zurek, ``Reduction of a wave packet in quantum {B}rownian motion," {\em Phys. Rev. D}, {\bf 40}, 1071-1094 (1989).

15. C.~Anastopoulos, ``Frequently asked questions about decoherence," quant-ph/0011123 (2001).

16. R.~Hanson, ``Drift-diffusion in mangled worlds quantum mechanics," quant-ph/0303114 (2003).

17. R.I. Sutherland, ``A suggestive way of deriving the quantum probability rule," {\em Found. Phys.}, {\bf 13}, 379-386 (2000).

\end{document}